\documentclass{article}
\usepackage{spconf,amsmath,graphicx,hyperref,multirow,booktabs,setspace}
\usepackage{makecell}

\ninept

\title{Summary of The Inaugural Music Source Restoration Challenge}

\name{
  \begin{tabular}{c}
  Yongyi Zang$^{1}$ \qquad
  Jiarui Hai$^{2}$ \qquad
  Wanying Ge$^{1}$ \qquad
  Qiuqiang Kong$^{3}$ \\
  Zheqi Dai$^{3}$ \qquad
  Helin Wang$^{2}$ \qquad
  Yuki Mitsufuji$^{4}$ \qquad
  Mark D.\ Plumbley$^{5}$
  \end{tabular}
}

\address{
    $^1$Independent Researcher \qquad
    $^2$Johns Hopkins University \qquad \\
    $^3$The Chinese University of Hong Kong \qquad
    $^4$Sony AI \qquad
    $^5$King's College London
}

\begin{document}
\maketitle

\begin{abstract}
Music Source Restoration (MSR) aims to recover original, unprocessed instrument stems from professionally mixed and degraded audio, requiring reversal of both production effects and real-world degradations. We present the inaugural MSR Challenge, featuring objective evaluation using Multi-Mel-SNR, Zimtohrli, and FAD-CLAP on studio-produced mixtures, alongside subjective evaluation on real-world degraded recordings. Five teams participated. The winning system achieved 4.46 dB Multi-Mel-SNR and 3.47 MOS-Overall, representing 91\% and 18\% relative improvements over the second-place system respectively. Per-stem analysis reveals that restoration difficulty varies substantially by instrument, with bass averaging 4.59 dB across all teams while percussion averages only 0.29 dB. The dataset, evaluation protocols, and baselines are available at \url{https://msrchallenge.com/}.
\end{abstract}

\begin{keywords}
Music Source Restoration, Audio Signal Processing, Deep Learning
\end{keywords}

\vspace{-0.3cm}
\section{Introduction}
\label{sec:intro}
\vspace{-0.15cm}

Music Source Separation (MSS) has traditionally assumed that mixtures are linear combinations of individual source signals~\cite{cano2018musical,vincent2006performance}. However, real-world recordings violate this assumption. During production, audio engineers apply equalization, dynamic range compression, reverberation, and mastering effects that alter the spectral and temporal characteristics of each instrument. During transmission and storage, recordings undergo further degradation through lossy audio codecs~\cite{defossezhigh,kumar2023high} and acoustic artifacts such as noise and distortion. Music Source Restoration (MSR)~\cite{zang2025music} extends MSS to address these complexities by recovering the original, unprocessed source signals from degraded mixtures.

The MSS community has established rigorous evaluation frameworks through competitions such as the Music Demixing Challenge~\cite{mitsufuji2022music} and the Sound Demixing Challenge~\cite{uhlich2023sound,solovyev2023benchmarks}, using datasets including MUSDB18-HQ~\cite{rafii2019musdb18} and MoisesDB~\cite{pereira2023moisesdb}. However, these benchmarks cannot evaluate restoration fidelity because their ground-truth stems already contain production effects applied during mixing. The MSR Challenge addresses this gap by providing the first benchmark with truly unprocessed reference stems, enabling evaluation of both separation accuracy and restoration quality.

This paper describes the challenge setup in Section~\ref{sec:setup}, presents results in Section~\ref{sec:results}, summarizes participating systems in Section~\ref{sec:systems}, and discusses key findings in Section~\ref{sec:discussion}.

\vspace{-0.3cm}
\section{Challenge Setup}
\label{sec:setup}
\vspace{-0.15cm}

\subsection{Task and Data}

The challenge task requires systems to restore eight instrument stems from mixed audio: vocals, guitars, keyboards, bass, synthesizers, drums, percussion, and orchestral elements. Given a degraded mixture as input, systems must output the original, unprocessed version of each stem before any production effects were applied.

The validation set, called MSRBench~\cite{zang2025msrbench}, contains 2,000 professionally mixed 10-second clips at 48~kHz stereo, with parallel unprocessed and processed stems for each clip. Mixtures are evaluated under 13 conditions: the original mastered audio plus 12 degradation types. These degradations span analog artifacts (radio transmission, cassette tape, vinyl records, live room acoustics), traditional lossy codecs (AAC and MP3 at 64 and 128 kbps)~\cite{tortosa2014optimal}, and neural audio codecs (DAC~\cite{kumar2023high} and Encodec~\cite{defossezhigh}).

The challenge includes two test sets. The non-blind test set contains 1,000 clips with ground-truth stems available, enabling computation of objective metrics. The blind test set contains 500 clips representing real-world degradation scenarios without ground truth: historical cylinder recordings from the early 1900s, live concert recordings, FM radio broadcasts, and low-bitrate streaming audio. This blind set tests generalization to degradations not seen during training.

\vspace{-0.15cm}
\subsection{Evaluation Metrics}

For objective evaluation on the non-blind test set, we employ three complementary metrics. Multi-Mel-SNR measures spectro-temporal reconstruction accuracy across multiple time-frequency resolutions, designed to avoid the phase oversensitivity of traditional waveform-domain metrics like SDR~\cite{vincent2006performance}. Zimtohrli~\cite{alakuijala2025zimtohrli} models perceptual similarity using psychoacoustic principles including gammatone filterbank analysis and temporal masking. FAD-CLAP~\cite{wu2023large} captures semantic similarity by computing Fr\'{e}chet distance over CLAP embeddings, measuring whether the restored audio sounds like the correct instrument.

For subjective evaluation on the blind test set, professional audio engineers rate each restored sample on three dimensions using 5-point Mean Opinion Scores (MOS). MOS-Separation measures how well the output isolates the target instrument from the mixture. MOS-Restoration assesses how effectively production effects and degradations have been removed. MOS-Overall captures the combined perceptual quality of the restored stem.

\vspace{-0.3cm}
\section{Results}
\label{sec:results}
\vspace{-0.15cm}

Five teams submitted results: xlancelab, CUPAudioGroup, AC\_DC, Hachimi, and cp-jku. Tables~\ref{tab:objective} and~\ref{tab:subjective} present the overall objective and subjective results, while Table~\ref{tab:perstem} provides per-stem Multi-Mel-SNR scores.

\begin{table}[t]
\centering
\vspace{-0.2cm}
\caption{Objective evaluation results on the non-blind test set. MMSNR: Multi-Mel-SNR in dB ($\uparrow$); Zimtohrli ($\downarrow$); FAD-CLAP ($\downarrow$).}
\vspace{0.1cm}
\small
\begin{tabular}{lccc}
\toprule
\textbf{Team} & \textbf{MMSNR} & \textbf{Zimt} & \textbf{FAD} \\
\midrule
xlancelab & 4.46 & 0.014 & 0.199 \\
CUPAudioGroup & 2.34 & 0.016 & 0.225 \\
AC\_DC & 1.45 & 0.018 & 0.291 \\
Hachimi & 2.00 & 0.018 & 0.294 \\
cp-jku & 0.83 & 0.019 & 0.381 \\
\bottomrule
\end{tabular}
\label{tab:objective}
\vspace{-0.3cm}
\end{table}

\begin{table}[t]
\centering
\caption{Subjective evaluation results on the blind test set (MOS on 1--5 scale, $\uparrow$).}
\vspace{0.1cm}
\small
\setlength{\tabcolsep}{3pt}
\begin{tabular}{lccc}
\toprule
\textbf{Team} & \textbf{Sep} & \textbf{Rest} & \textbf{Overall} \\
\midrule
xlancelab & 4.24 & 3.39 & 3.47 \\
CUPAudioGroup & 3.84 & 2.92 & 2.93 \\
Hachimi & 3.58 & 2.63 & 2.72 \\
AC\_DC & 3.54 & 2.48 & 2.54 \\
cp-jku & 3.55 & 2.08 & 2.14 \\
\bottomrule
\end{tabular}
\label{tab:subjective}
\vspace{-0.3cm}
\end{table}

\begin{table}[t]
\centering
\caption{Per-stem Multi-Mel-SNR in dB ($\uparrow$).}
\vspace{0.1cm}
\small
\setlength{\tabcolsep}{3pt}
\begin{tabular}{lcccccccc}
\toprule
\textbf{Team} & \textbf{Voc} & \textbf{Gtr} & \textbf{Key} & \textbf{Syn} & \textbf{Bass} & \textbf{Drm} & \textbf{Prc} & \textbf{Orc} \\
\midrule
xlancelab & 1.56 & 3.95 & 6.71 & 2.26 & 8.22 & 5.65 & 1.17 & 6.17 \\
CUPAudio & 1.36 & 1.95 & 2.76 & 0.98 & 5.29 & 4.92 & 0.16 & 1.32 \\
AC\_DC & 1.05 & 1.14 & 1.82 & 0.95 & 2.86 & 2.73 & 0.02 & 1.05 \\
Hachimi & 1.05 & 1.03 & 3.34 & 1.24 & 5.05 & 2.85 & 0.00 & 1.45 \\
cp-jku & 0.84 & 1.29 & 0.78 & 0.64 & 1.55 & 0.96 & 0.10 & 0.51 \\
\midrule
\textit{Average} & \textit{1.17} & \textit{1.87} & \textit{3.08} & \textit{1.21} & \textit{4.59} & \textit{3.42} & \textit{0.29} & \textit{2.10} \\
\bottomrule
\end{tabular}
\label{tab:perstem}
\vspace{-0.3cm}
\end{table}

The xlancelab team ranked first across all metrics, achieving 91\% relative improvement in Multi-Mel-SNR and 18\% in MOS-Overall compared to the second-place CUPAudioGroup. The objective and subjective rankings show strong agreement (Spearman $\rho = 0.9$), with only AC\_DC and Hachimi swapping positions between the two evaluation paths.

The per-stem results in Table~\ref{tab:perstem} reveal that restoration difficulty varies substantially across instrument types. Averaging across all teams, bass achieves the highest scores at 4.59 dB, followed by drums (3.42 dB) and keyboards (3.08 dB). Percussion proves consistently challenging, with an average of only 0.29 dB and four of five teams scoring below 0.2 dB. Notably, vocals average only 1.17 dB despite being a primary focus in traditional MSS research~\cite{stoller2018wave,jansson2017singing}. The xlancelab system shows its largest advantages on polyphonic sources, gaining 4.72 dB on orchestral elements and 3.37 dB on keyboards over the second-place team, while the advantage narrows to just 0.20 dB on vocals.

\vspace{-0.3cm}
\section{Participating Systems}
\label{sec:systems}
\vspace{-0.15cm}

All participating teams built upon transformer-based architectures that have proven effective for music source separation. We briefly describe each system below.

The xlancelab team employed sequential BSRoformers~\cite{lu2024music,wang2023mel}, a band-split transformer architecture, with three pretrained modules applied in sequence: separation, dereverberation, and denoising. Their training used L1 loss combined with multi-resolution STFT loss on MoisesDB~\cite{pereira2023moisesdb} and a manually cleaned version of the RawStems dataset~\cite{zang2025music}.

CUPAudioGroup built an ensemble combining three complementary architectures: BSRNN~\cite{luo2023music} (band-split recurrent neural network), BSRoformer~\cite{lu2024music}, and MDX23. All models were initialized from pretrained weights and trained on RawStems, MUSDB18-HQ~\cite{rafii2019musdb18}, and MoisesDB.

The AC\_DC team proposed DTT-BSR, a novel architecture combining DTTNet (a dual-path TFC-TDF U-Net)~\cite{chen2024music} with Band-Sequence Modeling~\cite{luo2023music} and a RoPE Transformer bottleneck. They employed adversarial training using a multi-frequency discriminator.

Hachimi adapted a mel-band separation backbone~\cite{wang2023mel} with combined reconstruction and GAN losses. They used the most diverse training data, combining six datasets: MUSDB18-HQ, MoisesDB, MedleyDB~\cite{bittner2014medleydb}, RawStems, URMP~\cite{li2018creating}, and MAESTRO~\cite{hawthorne2018enabling}.

The cp-jku team used BSRoformer~\cite{lu2024music} for separation and HiFi++ GAN bundle (SpectralUNet, Upsampler, WaveUNet, SpectralMaskNet) for restoration, training eight source-specific expert models with LoRA adapters.

\vspace{-0.3cm}
\section{Discussion}
\label{sec:discussion}
\vspace{-0.15cm}

\noindent \textbf{Multi-stage processing benefits top systems.} The top two systems adopted multi-stage processing approaches rather than attempting to solve restoration in a single model. The xlancelab system chains separation, dereverberation, and denoising modules sequentially, while CUPAudioGroup ensembles three complementary separation models. This modularity enables leveraging pretrained MSS checkpoints~\cite{solovyev2023benchmarks} and reduces the complexity that each processing stage must handle. However, multi-stage processing alone does not guarantee success: cp-jku also used a two-stage pipeline but ranked fifth, suggesting that architectural choices within each stage remain critical.

\noindent \textbf{Data quality matters more than quantity.} All participating teams used the RawStems dataset~\cite{zang2025music} for training, but only xlancelab invested effort in manually cleaning it to address known alignment and source leakage issues in the original data. Despite Hachimi combining six different datasets compared to xlancelab's two, xlancelab achieved 91\% higher Multi-Mel-SNR, suggesting that data quality is more important than data diversity for this task.

\noindent \textbf{Simple reconstruction losses outperform adversarial training.} The top two teams relied exclusively on L1 and STFT reconstruction losses without adversarial training. In contrast, teams employing GAN-based training (AC\_DC, Hachimi, and cp-jku) ranked third through fifth. This suggests that adversarial training may not provide clear benefits for MSR, or that it requires particularly careful tuning that these systems did not achieve.

\noindent \textbf{Polyphonic and transient sources pose distinct challenges.} The xlancelab system's performance advantage concentrates on polyphonic sources such as orchestral elements (+4.72 dB over second place) and keyboards (+3.37 dB), where complex harmonic relationships must be preserved. However, the advantage narrows substantially for monophonic sources like vocals (+0.20 dB). Meanwhile, percussion remains difficult for all systems (average 0.29 dB, compared to 4.59 dB for bass) due to its impulsive, broadband nature, which poses fundamental challenges for phase reconstruction.

\vspace{-0.3cm}
\section{Conclusion}
\label{sec:conclusion}
\vspace{-0.15cm}

The MSR Challenge has established the first standardized benchmark for music source restoration. The results demonstrate that sequential and ensemble architectures leveraging pretrained MSS models with simple reconstruction losses achieve the best performance. The 16$\times$ performance gap between bass (4.59 dB) and percussion (0.29 dB) averaged across all teams indicates that source-specific approaches may be necessary for practical deployment. The dataset and baseline implementations are publicly available at \url{https://msrchallenge.com/}.

\vspace{-0.3cm}
\section{Acknowledgements}
\label{sec:acknowledgement}
\vspace{-0.15cm}
This work was supported by the Engineering and Physical Sciences Research Council (EPSRC) under grant numbers EP/T019751/1, EP/Y028805/1, and UKRI397. For the purpose of open access, the authors have applied a Creative Commons Attribution (CC BY) licence to any Author Accepted Manuscript version arising.

\bibliographystyle{IEEEbib}
\bibliography{strings,refs}

\end{document}